      [1999/12/01 v1.4c Il Nuovo Cimento]
\documentclass{cimento}

\usepackage{pstricks,pst-node,pst-plot,pst-3d,pst-3dplot,pst-text,pst-eps}
\usepackage{amsmath, amsthm, amssymb, bm}
\usepackage{verbatim} 
\usepackage{graphicx}
\usepackage{subfigure}

\title
      {Casimir interaction energies for magneto-electric $\delta$-function 
plates.}
\author{Kimball A.~Milton\from{ins:y}\ETC,
        Prachi Parashar\from{ins:y},
        Martin Schaden\from{ins:z},
\atque
        K.~V.~Shajesh\from{ins:x}\\
    }
\instlist{
\inst{ins:y} Homer L.~Dodge Department of Physics and Astronomy, 
             University of Oklahoma, Norman, Oklahoma 73019, USA
\inst{ins:z} Department of Physics, Rutgers, The State University 
             of New Jersey, Newark, New Jersey 07102, USA
\inst{ins:x} Department of Materials Science and Engineering, 
             Iowa State University of Science and Technology, 
             Ames, Iowa 50011, USA}
\PACSes{\PACSit{03.70.+k}{11.80.La. -- 34.35.+a -- 42.50.Lc}}
\begin{document}

\maketitle

\begin{abstract}
We present boundary conditions for the electromagnetic fields on a 
$\delta$-function plate, having both electric and magnetic properties, sandwiched 
between two magneto-electric semi-infinite half spaces. The optical properties 
for an isolated $\delta$-function plate are shown to be independent of the 
longitudinal material properties of the plate. The Casimir-Polder energy between 
an isotropically polarizable atom and a magneto-electric $\delta$-function plate 
is attractive for a purely electric $\delta$-function plate, repulsive for a 
purely magnetic $\delta$-function plate, and vanishes for the simultaneous 
perfect conductor limit of both electric and magnetic properties of the 
$\delta$-function plate. The interaction energy between two identical 
$\delta$-function plates is always attractive. It can be attractive or repulsive 
when the plates have electric and magnetic properties interchanged and reproduces 
Boyer's result for the interaction energy between perfectly conducting electric 
and magnetic plates.
The change in the Casimir-Polder energy in the presence of a $\delta$-function 
plate on a magneto-electric substrate is substantial when the substrate is a 
weak dielectric. 
\end{abstract}

\section{Introduction}
Infinitesimally thin perfectly conducting surfaces have often been used at least 
since the first rigorous exact solution of diffraction of a plane wave by a 
half-plate of infinitesimal thickness given by Sommerfeld in 
1896~\cite{Sommerfeld:1896,Sommerfeld:2004}. 
Another iconic example, in the field of Casimir physics, is Boyer's calculation 
of the repulsive Casimir pressure for such an infinitesimally thin perfectly 
conducting spherical shell~\cite{Boyer:1968uf}. 
A closed perfectly conducting infinitesimally thin surface is like an ``electric 
wall'' that decouples two regions of space~\cite{Milton:2006}. Therefore, it is 
sufficient to consider only the region of interest where the interaction is 
occurring. However, examples like an infinitesimally thin half plane or a 
perfectly conducting plate with an aperture~\cite{Johnson-PhysRevLett.105.090403} 
require the consideration of the other side of the perfectly conducting surface. 

Boundary conditions on an electric material of infinitesimal thickness were first 
derived by Barton in Refs.~\cite{Barton:2004sph,Barton:2004sp2,Barton:2004sp3} 
who observed that an infinitesimally thin conducting surface imposes non-trivial 
boundary conditions on the electromagnetic fields and in 
Refs.~\cite{Barton:2005eps,Barton:2005fst} considered ``a fluid model of an 
infinitesimally thin plasma sheet''. These boundary conditions were generalized 
for magneto-electric materials in Ref.~\cite{Parashar-PhysRevD.86.085021}.

References~\cite{Bordag:1985th,Bordag:1992cp} were the first to use a 
$\delta$-function potential to mathematically represent an infinitesimally thin 
surface. Robaschik and Wieczorek in~\cite{Robaschik:1994bc} proposed that two 
different boundary conditions could be satisfied on a perfectly conducting 
electric $\delta$-function plate, and Bordag in~\cite{Bordag:2004smr} further 
claimed that the interaction energy between an atom and a $\delta$-function plate 
satisfying these two boundary conditions are not identical.
These confusions were discussed in detail and resolved 
in~\cite{Parashar-PhysRevD.86.085021} in which we showed that the electric 
Green's dyadic obtained using both boundary conditions were identical and 
therefore corresponded to the same physical situation.

In~\cite{Parashar-PhysRevD.86.085021} we derived the boundary conditions on a 
$\delta$-function plate having both electric and magnetic properties, which will 
be termed as a magneto-electric $\delta$-function plate in this paper, and showed 
that such a plate can be realized physically in the so-called ``thin-plate'' 
limit. We presented results for the interaction energy between two such 
$\delta$-function plates and between an atom and a $\delta$-function plate when 
they have purely electric properties. 

In this paper we study examples involving magneto-electric $\delta$-function 
plates. In the following section, we briefly present the derivation of the 
boundary conditions on an infinitesimally thin magneto-electric $\delta$-function 
plate sandwiched between two magneto-electric semi-infinite half spaces and 
present solutions for the magnetic and electric Green's functions. In the 
thin-plate limit, $\zeta d \ll \sqrt{\zeta_p d} \ll 1$, where 
$\omega_p^2=\zeta_p/d$ is the plasma frequency of the material, a vanishing 
thickness $d$ of the plate reproduces the optical properties of an electric 
$\delta$-function plate. 
The suggestion is that a theoretical calculation, for example, for a corrugated 
surface, could be greatly simplified if the boundaries in consideration could be 
approximated by their respective $\delta$-function forms. 

In the subsequent sections we consider the change in the Casimir-Polder energy 
due to the presence of a magneto-electric $\delta$-function plate on the surface 
of a magneto-electric semi-infinite half space and the Casimir interaction energy 
between two magneto-electric $\delta$-function plates. In experiments thin films 
of metals are grown on a substrate and their properties are not continuous, 
transitioning from insulator to metal abruptly. This is in contrast to the 
$\delta$-function plate, which has continuous properties, and it is therefore not 
clear how to compare our results with experimentally realizable thin metal films. 
On the other hand, it is well known that a coat of a thin dielectric film on a 
metal surface changes the reflectivity of the metal surface. Thus in principle, 
one could think of varying the Casimir interaction energy between two surfaces by 
coating them with the physically realizable $\delta$-function plates discussed in 
Sec.~\ref{phys-delta}.


\section{Boundary conditions on an infinitesimally thin magneto-electric 
$\delta$-function plate}
\label{delta-def}
\begin{figure}\begin{center} 
\includegraphics[width=6.5cm]{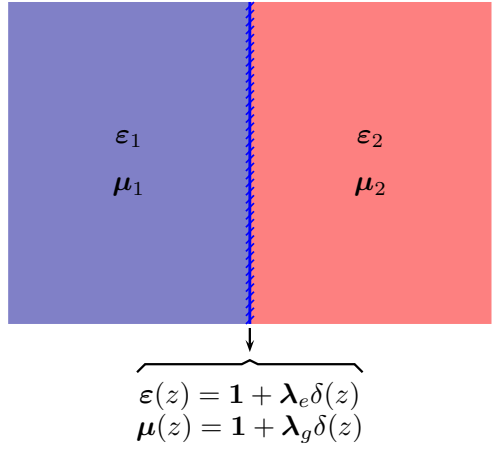}
\caption{A magneto-electric $\delta$-function plate sandwiched between two 
magneto-electric semi-infinite slabs.}
\label{slabs-delta-fig}
\end{center}\end{figure}

We consider a magneto-electric $\delta$-function plate sandwiched between two 
uniaxial magneto-electric materials as shown in Fig.~\ref{slabs-delta-fig}. 
The electric permittivity $\varepsilon$ and the magnetic permeability $\mu$ 
for this system are described by
\begin{equation}
\bm{\varepsilon} (z) = \varepsilon^\perp (z)\, {\bf 1}_\perp 
+ \varepsilon^{||}(z) \, \hat{\bf z} \,\hat{\bf z} 
\qquad \text{and} \qquad
\bm{\mu} (z) = \mu^\perp (z)\, {\bf 1}_\perp 
+ \mu^{||} (z) \,\hat{\bf z} \,\hat{\bf z},
\label{epmu-ds}%
\end{equation}
where $z=a$ is the position of the interface, and 
\begin{subequations}
\begin{eqnarray}
\varepsilon^{\perp,||} (z) &=& 1 + (\varepsilon^{\perp,||}_1-1) \theta(a-z) 
+ (\varepsilon^{\perp,||}_2-1) \theta(z-a) + \lambda^{\perp,||}_e \delta(z-a),
\\
\mu^{\perp,||} (z) &=& 1 + (\mu^{\perp,||}_1-1) \theta(a-z) 
+ (\mu^{\perp,||}_2-1) \theta(z-a) + \lambda^{\perp,||}_g \delta(z-a).
\end{eqnarray}%
\label{eps-def}%
\end{subequations}
The electric permittivity and magnetic permeability are in general frequency 
dependent. The Maxwell equations in the absence of charges and currents, in 
frequency space, are  
\begin{equation}
{\bm \nabla} \times {\bf E} = i \omega {\bf B}
\qquad \text{and} \qquad 
-{\bm \nabla} \times {\bf H} = i \omega ({\bf D} + {\bf P}), 
\label{ME-cross}%
\end{equation}
where we assume the fields ${\bf D}$ and ${\bf B}$ are linearly dependent 
on the electric and magnetic fields ${\bf E}$ and ${\bf H}$ as
\begin{equation}
{\bf D}({\bf x},\omega) 
= {\bm \varepsilon}({\bf x};\omega) \cdot {\bf E}({\bf x},\omega)
\qquad \text{and} \qquad
{\bf B}({\bf x},\omega) 
= {\bm \mu}({\bf x};\omega) \cdot {\bf H}({\bf x},\omega),
\label{DB=emuEB}%
\end{equation} 
and ${\bf P}$ is an external source of polarization. 
\subsection{Boundary conditions}
The Maxwell equations in Eq.~(\ref{ME-cross}) decouple into transverse 
electric (TE) and transverse magnetic (TM) modes for planar geometries. The 
boundary conditions on the electric and magnetic fields ${\bf E}$ and ${\bf H}$ 
are obtained by integrating across the $\delta$-function boundary. We get 
additional contributions to the standard boundary conditions at the interface of 
two media due to the presence of the magneto-electric $\delta$-function plate 
as follows:
\begin{subequations}
\begin{align}
&\text{\underline{TM}}&\text{\underline{TE}}&\nonumber\\
E_1\Big|^{z=a+}_{z=a-} &= i\omega \lambda_g^\perp H_2(a),
& H_1\Big|^{z=a+}_{z=a-} &= -i\omega \lambda_e^\perp E_2(a),\\
H_2\Big|^{z=a+}_{z=a-} &=i\omega \lambda_e^\perp E_1(a),
& E_2\Big|^{z=a+}_{z=a-} &= -i \omega \lambda_g^\perp H_1(a),\\
D_3\Big|^{z=a+}_{z=a-} &= -ik_\perp \lambda_e^\perp E_1(a),
& B_3\Big|^{z=a+}_{z=a-} &= -ik_\perp \lambda_g^\perp H_1(a) \label{TM-bc-D3}.
\end{align}
\end{subequations}
In addition we get the constraints,
\begin{equation}
\lambda_e^{||} E_3(a) = 0 
\qquad \text{and} \qquad
\lambda_g^{||} H_3(a) = 0,
\label{constraint}%
\end{equation}
which imply that optical properties of the magneto-electric $\delta$-function 
plate are necessarily anisotropic unless $E_3(a)=0$ and $H_3(a)=0$. These 
restrictions are implicit in the model considered by Barton~\cite{Barton:2005eps}.
\subsection{Green's functions}
We use the Green's function technique to obtain the electric and magnetic fields 
${\bf E}$ and ${\bf H}$: 
\begin{equation}
{\bf E}({\bf x}) 
= \int d^3x^\prime \,{\bm \Gamma} ({\bf x},{\bf x}^\prime) \cdot 
  {\bf P}({\bf x}^\prime)
\qquad \text{and} \qquad
{\bf H}({\bf x}) 
= \int d^3x^\prime \,{\bm \Phi} ({\bf x},{\bf x}^\prime) \cdot 
  {\bf P}({\bf x}^\prime),
\label{E=Gp,H=PP}%
\end{equation}
in terms of the electric Green's dyadic ${\bm \Gamma} ({\bf x},{\bf x}^\prime)$ 
and magnetic Green's dyadic ${\bm \Phi} ({\bf x},{\bf x}^\prime)$ respectively. 
Using translational symmetry we can Fourier transform the Green's dyadics in the 
$xy$-plane, for example,
\begin{equation}
{\bm\Gamma}({\bf x},{\bf x}^\prime;\omega)
= \int \frac{d^2k_\perp}{(2\pi)^2} 
\,e^{i{\bf k}_\perp \cdot ({\bf x}-{\bf x}^\prime)_\perp} 
{\bm\gamma}(z,z^\prime;{\bf k}_\perp,\omega).
\end{equation}
The reduced Green's dyadics ${\bm \gamma} (z,z^\prime)$ and 
${\bm \phi} (z,z^\prime)$, in the coordinate system where ${\bm k}_\perp $ lies 
in x direction, are
\begin{equation}
{\bm\gamma}
= \left[ \begin{array}{ccc}
\frac{1}{\varepsilon^\perp} \frac{\partial}{\partial z}
\frac{1}{\varepsilon^{\prime\perp}} \frac{\partial}{\partial z^\prime}
 g^H & 0 &
\frac{1}{\varepsilon^\perp} \frac{\partial}{\partial z}
\frac{ik_\perp}{\varepsilon^{\prime||}} g^H \\[2mm]
0 & \omega^2 g^E & 0 \\[2mm]
-\frac{ik_\perp}{\varepsilon^{||}} \frac{1}{\varepsilon^{\prime\perp}} 
\frac{\partial}{\partial z^\prime} g^H & 0 &
-\frac{ik_\perp}{\varepsilon^{||}}
\frac{ik_\perp}{\varepsilon^{\prime||}} g^H
\end{array} \right]
\label{Gamma=gE}
\end{equation}
and
\begin{equation}
{\bm\phi}
= i\omega \left[ \begin{array}{ccc}
0 & \frac{1}{\mu^\perp} \frac{\partial}{\partial z} g^E
& 0 \\[2mm]
\frac{1}{\varepsilon^{\prime\perp}}
\frac{\partial}{\partial z^\prime} g^H & 0 &
\frac{ik_\perp}{\varepsilon^{\prime||}} g^H \\[2mm]
0 & -\frac{ik_\perp}{\mu^{||}} g^E & 0
\end{array} \right],
\label{Phi=gH}
\end{equation}
where we have suppressed the $z$ and $z^\prime$ dependence and 
$\varepsilon^\prime$ is evaluated at point $z^\prime$. In Eq.~(\ref{Gamma=gE}) 
we have omitted a contact term involving $\delta(z-z^\prime)$, which does not 
contribute to interaction energies between disjoint objects. The magnetic 
Green's function $g^H(z,z^\prime)$ and the electric Green's function 
$g^E(z,z^\prime)$ satisfy 
\begin{subequations}
\begin{eqnarray}
\left[ - \frac{\partial}{\partial z} \frac{1}{\varepsilon^\perp(z)}
\frac{\partial}{\partial z} + \frac{k_\perp^2}{\varepsilon^{||}(z)} 
-\omega^2 \mu^\perp(z) \right] g^H(z,z^\prime) &=& \delta(z-z^\prime), 
\label{greenH} \\
\left[ - \frac{\partial}{\partial z} \frac{1}{\mu^\perp(z)}
\frac{\partial}{\partial z} + \frac{k_\perp^2}{\mu^{||}(z)} 
-\omega^2 \varepsilon^\perp(z) \right] g^E(z,z^\prime) &=& \delta(z-z^\prime), 
\label{greenE}
\end{eqnarray}%
\label{green-funs}%
\end{subequations}
where the material properties $\varepsilon^\perp(z)$ and $\mu^\perp(z)$ are 
given by Eqs.~(\ref{eps-def}). We obtain the boundary conditions on the magnetic 
Green's functions using Eq.~(\ref{TM-bc-D3}) for TM mode,
\begin{subequations}
\begin{eqnarray}
g^H \Big|^{z=a+}_{z=a-}
&=& \lambda^\perp_e \frac{1}{\varepsilon^\perp} \partial_z
g^H \bigg|_{z=a} , \\
\frac{1}{\varepsilon^\perp} \partial_z
g^H \bigg|^{z=a+}_{z=a-}
&=& \zeta^2 \lambda^\perp_g g^H\Big|_{z=a}.
\end{eqnarray}%
\label{gH-bc}%
\end{subequations} 
Similarly, using Eq.~(\ref{TM-bc-D3}) for TE mode, the boundary 
conditions on the electric Green's function are 
\begin{subequations}
\begin{eqnarray}
g^E \Big|^{z=a+}_{z=a-}
&=& \lambda^\perp_g \frac{1}{\mu^\perp} \partial_z
g^E \bigg|_{z=a}, \\
\frac{1}{\mu^\perp} \partial_z
g^E \bigg|^{z=a+}_{z=a-}
&=& \zeta^2 \lambda^\perp_e g^E\Big|_{z=a}.
\end{eqnarray}%
\label{gE-bc}%
\end{subequations} 
Here $\zeta$ is the imaginary frequency obtained after a Euclidean rotation. 
We evaluate quantities that are discontinuous on the magneto-electric 
$\delta$-function plate using the averaging prescription described 
in~\cite{CaveroPelaez:2008tj}.

The solution for the magnetic Green's function satisfying the boundary conditions 
in Eq.~(\ref{gH-bc}) is
\begin{equation}
g^H(z,z^\prime) = 
\begin{cases}
\frac{1}{2\bar{\kappa}^H_1} \Big[ e^{-\kappa^H_1 |z-z^\prime|}
+ r^H_{12} \, e^{-\kappa^H_1 |z-a|} e^{-\kappa^H_1 |z^\prime-a|} \Big], 
& \text{if} \quad z,z^\prime < a, \\[2mm]
\frac{1}{2\bar{\kappa}^H_2} \Big[ e^{-\kappa^H_2 |z-z^\prime|}
+ r^H_{21} \, e^{-\kappa^H_2 |z-a|} e^{-\kappa^H_2 |z^\prime-a|} \Big], 
& \text{if} \quad a < z,z^\prime, \\[2mm]
\frac{1}{2\bar{\kappa}^H_2} \, t^H_{21} \, 
e^{-\kappa^H_1 |z-a|} e^{-\kappa^H_2 |z^\prime-a|}, 
& \text{if} \quad z<a < z^\prime, \\[2mm]
\frac{1}{2\bar{\kappa}^H_1} \,t^H_{12} \,
e^{-\kappa^H_2 |z-a|} e^{-\kappa^H_1 |z^\prime-a|}, 
& \text{if} \quad z^\prime <a <z, 
\end{cases}
\label{gH-sol}
\end{equation}
where the reflection and transmission coefficients are
\begin{subequations}
\begin{eqnarray}
r^H_{ij} =
\frac{\bar{\kappa}^H_i 
\Big( 1 + \frac{\lambda^\perp_e \bar{\kappa}^H_j}{2} \Big)
\Big( 1 - \frac{\lambda^\perp_g \zeta^2}{2\bar{\kappa}^H_i} \Big)
-\bar{\kappa}^H_j 
\Big( 1 - \frac{\lambda^\perp_e \bar{\kappa}^H_i}{2} \Big)
\Big( 1 + \frac{\lambda^\perp_g \zeta^2}{2\bar{\kappa}^H_j} \Big) }
{ \bar{\kappa}^H_i 
\Big( 1 + \frac{\lambda^\perp_e \bar{\kappa}^H_j}{2} \Big)
\Big( 1 + \frac{\lambda^\perp_g \zeta^2}{2\bar{\kappa}^H_i} \Big)
+\bar{\kappa}^H_j 
\Big( 1 + \frac{\lambda^\perp_e \bar{\kappa}^H_i}{2} \Big)
\Big( 1 + \frac{\lambda^\perp_g \zeta^2}{2\bar{\kappa}^H_j} \Big) },\\
t^H_{ij} =
\frac{\bar{\kappa}^H_i 
\Big( 1 + \frac{\lambda^\perp_e \bar{\kappa}^H_i}{2} \Big)
\Big( 1 - \frac{\lambda^\perp_g \zeta^2}{2\bar{\kappa}^H_i} \Big)
+\bar{\kappa}^H_i 
\Big( 1 - \frac{\lambda^\perp_e \bar{\kappa}^H_i}{2} \Big)
\Big( 1 + \frac{\lambda^\perp_g \zeta^2}{2\bar{\kappa}^H_i} \Big) }
{ \bar{\kappa}^H_i 
\Big( 1 + \frac{\lambda^\perp_e \bar{\kappa}^H_j}{2} \Big)
\Big( 1 + \frac{\lambda^\perp_g \zeta^2}{2\bar{\kappa}^H_i} \Big)
+\bar{\kappa}^H_j 
\Big( 1 + \frac{\lambda^\perp_e \bar{\kappa}^H_i}{2} \Big)
\Big( 1 + \frac{\lambda^\perp_g \zeta^2}{2\bar{\kappa}^H_j} \Big) },
\end{eqnarray}%
\label{rH-tH}%
\end{subequations} 
with
\begin{equation}
\kappa_i^H
= \sqrt{k_\perp^2 \frac{\varepsilon^\perp_i}{\varepsilon^{||}_i}
+ \zeta^2 \varepsilon^\perp_i \mu^\perp_i}
\qquad \text{and} \qquad
\bar{\kappa}^H_i = \frac{\kappa^H_i}{\varepsilon^\perp_i} 
=\sqrt{ \frac{k_\perp^2}{\varepsilon^\perp_i \varepsilon^{||}_i}
+ \zeta^2 \frac{\mu^\perp_i}{\varepsilon^\perp_i} }.
\label{kaiH}
\end{equation}
The electric Green's function is obtained by replacing 
${\bm \varepsilon} \leftrightarrow {\bm \mu}$ and $H\to E$ everywhere.
%
Notice that the reflection and transmission coefficients are independent of 
$\lambda^{||}_e$ and  $\lambda^{||}_g$, which implies that the optical properties 
of the magneto-electric $\delta$-function plates are independent of the 
longitudinal components of the material properties.
\subsection{Green's function for an isolated magneto-electric $\delta$-function 
plate in vacuum}
\label{single-plate}
Green's function for a magneto-electric $\delta$-function plate in vacuum is 
obtained by setting $\varepsilon^\perp_i=\varepsilon^{||}_i=1$ and 
$\mu^\perp_i=\mu^{||}_i=1$ in Eq.~(\ref{rH-tH}). The magnetic Green's function 
in compact form is 
\begin{equation}
g^H(z,z^\prime) = \frac{1}{2\kappa} e^{-\kappa |z-z^\prime|}
+ \big[ r^H_g + \text{sgn}(z-a) \text{sgn}(z^\prime -a) \, r^H_e \big]
\frac{1}{2\kappa} e^{-\kappa |z-a|} e^{-\kappa |z^\prime -a|},
\label{gHtp}
\end{equation}
where $\kappa^2 = k_\perp^2 + \zeta^2$. The electric and magnetic reflection 
coefficients are 
\begin{equation}
r^H_e = \frac{\lambda^\perp_e}{\lambda^\perp_e + \frac{2}{\kappa}},
\qquad \text{and} \qquad
r^H_g = - \frac{\lambda^\perp_g}{\lambda^\perp_g + \frac{2\kappa}{\zeta^2}},
\label{rHeg}
\end{equation}
which are defined for the cases $\lambda_g^\perp$ and $\lambda_e^\perp$ being 
zero, respectively. The total reflection coefficient for the magnetic mode is 
$r^H=r^H_g+r^H_e$. 
The TE reflection coefficient $r^E$ is obtained by replacing 
$e \leftrightarrow g$ and $H \to E$ in Eq.~(\ref{rHeg}). The TM and TE reflection 
coefficients vanish when simultaneously $\lambda_e\to\infty$ and 
$\lambda_g\to\infty$: The plate behaves like a perfect electric and perfect 
magnetic conductor, which we will refer as perfect magneto-electric conductor. 
This implies that a perfectly conducting magneto-electric $\delta$-function plate 
becomes transparent to the electromagnetic fields. 
\section{Physical realization of an electric $\delta$-function plate: Thin plate 
limit}
\label{phys-delta}
The $\delta$-function potential used to describe a magneto-electric plate in 
Sec.~\ref{delta-def} is a mathematical tool, which gives calculational 
ease. In case of a perfect conductor a $\delta$-function potential still 
serves as an accurate description of the physical system because the 
perfect conductor decouples the two regions in space. However, to describe 
a thin dielectric material slab of thickness $d$ using a $\delta$-function 
potential we need to use approximations on the material properties in the 
limit $d \to 0$. We can write a $\delta$-function as difference of two step functions describing a slab of thickness $d$ and take the limit $d \to 0$ 
after dividing out the thickness. Multiplying this construction by 
${\bm \lambda_e}$, we can read off the susceptibility of the slab 
as ${\bm \lambda_e}/d$.

The transverse magnetic and transverse electric reflection coefficient of 
a material slab of thickness $d$ is 
\begin{equation}
r_\text{thick}^H 
= -\frac{\left(\frac{\bar{\kappa}^H -\kappa}{\bar{\kappa}^H+\kappa} \right)
(1- e^{-2\kappa^Hd})} 
{\left[ 1- \left(\frac{\bar{\kappa}^H -\kappa}{\bar{\kappa}^H+\kappa} \right)^2
e^{-2\kappa^Hd}\right]} 
\qquad \text{and} \qquad
r_\text{thick}^E 
= -
\frac{\left(\frac{\kappa^E -\kappa}{\kappa^E+\kappa} \right)
(1- e^{-2\kappa^Ed})} 
{\left[ 1- \left(\frac{\kappa_i^E -\kappa}{\kappa^E+\kappa} \right)^2
e^{-2\kappa^Ed}\right]}. 
\label{thick-rs}%
\end{equation}
Naively taking the $d \to 0$ limit yields vanishing reflection coefficients. 
However, in the thin-plate limit, 
\begin{equation}
\zeta^2 \ll \frac{\zeta_p}{d} \ll \frac{1}{d^2}, 
\qquad \text{and} \qquad
k_\perp^2 \ll \frac{\zeta_p}{d} \ll \frac{1}{d^2}, 
\label{dtp-con}
\end{equation}
where $\zeta_p=\omega_p^2 d$ is the characteristic wave number of the material, 
the reflection coefficients for TM- and TE-modes exactly reproduce the 
reflection coefficients for a purely electric $\delta$-function plate: 
%
\begin{equation}
r_\text{thick}^H 
\xrightarrow[k_\perp d \ll \sqrt{\zeta_p d} \ll 1]
{\zeta d \ll \sqrt{\zeta_p d} \ll 1} 
r_e^H = \frac{\lambda_e^\perp}
{\lambda_e^\perp+\frac{2}{\kappa}}, 
\quad \text{and} \quad
r_\text{thick}^E 
\xrightarrow[k_\perp d \ll \sqrt{\zeta_p d} \ll 1]
{\zeta d \ll \sqrt{\zeta_p d} \ll 1}
r_e^E = -\frac{\lambda_e^\perp}
{\lambda_e^\perp+\frac{2\kappa}{\zeta^2}}. 
\label{thick-rs}%
\end{equation}%
%
It is worth noting that the reflection coefficients for both a thick slab and a 
$\delta$-function plate give same value in the perfect conductor limit, i.e., 
when the electrical permittivity goes to infinity. 
\section{Interaction energy between an electrically polarizable atom and a 
magneto-electric $\delta$-function plate}
\label{CP-energy}
In this section we consider the interaction of an atom with anisotropic 
electric polarizability 
${\bm\alpha} =\text{diag}(\alpha^\perp,\alpha^\perp, \alpha^{||})$ with a 
magneto-electric $\delta$-function plate. 
\subsection{Atom interacting with a magneto-electric $\delta$-function 
plate in vacuum}
\begin{figure}
\subfigure[Magneto-electric $\delta$-function plate]
{\includegraphics[width=48mm]{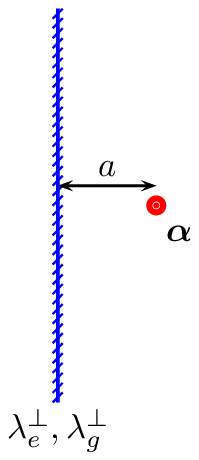}
\label{atom-tp-fig}}
\hspace{20mm}
\subfigure[Magneto-electric $\delta$-function plate on a dielectric slab]
{\includegraphics[width=48mm]{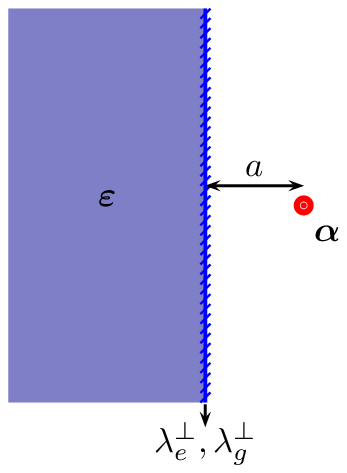}
\label{atom-plate-slab}}
\caption{Anisotropic atom in front of \subref{atom-tp-fig} a magneto-electric 
$\delta$-function plate, versus \subref{atom-plate-slab} a magneto-electric 
$\delta$-function plate on an anisotropic dielectric slab.}
\end{figure}

For the first case let us assume that the magneto-electric $\delta$-function 
plate is a stand-alone plate interacting with an electrically polarizable atom 
separated by a distance $a$, as shown in Fig.~\ref{atom-tp-fig}.
%
The Casimir-Polder energy between an anisotropic atom and a magneto-electric 
$\delta$-function plate for this case evaluates to 
\begin{equation}
E_{\delta\text{-atom}} =- 2\pi \int_{-\infty}^\infty \frac{d\zeta}{2\pi}
\int \frac{d^2k_\perp}{(2\pi)^2} \,\frac{e^{-2\kappa a}}{2\kappa}
\Big[ \alpha^\perp (\kappa^2 r^H -\zeta^2 r^E)
+ \alpha^{||} k_\perp^2 \,r^H \Big],
\label{CP12-tp}
\end{equation} 
where the TM and TE reflection coefficients for a $\delta$-function plate are 
provided in Sec.~\ref{single-plate}. More specifically, the TM reflection 
coefficient 
\begin{equation}
r^H=\frac{\lambda^\perp_{e}}{\lambda^\perp_{e} + \frac{2}{\kappa}}
 - \frac{\lambda^\perp_{g}}{\lambda^\perp_{g} + \frac{2\kappa}{\zeta^2}}
\label{rH-tot}
\end{equation}
and TE reflection coefficient $r^E$ is obtained by replacing 
$e \leftrightarrow g$ and $H \to E$ in Eq.~(\ref{rH-tot}). In the retarded limit 
we replace atomic polarizabilities by their static limits. 
\begin{figure}\begin{center}
\includegraphics[width=70mm]{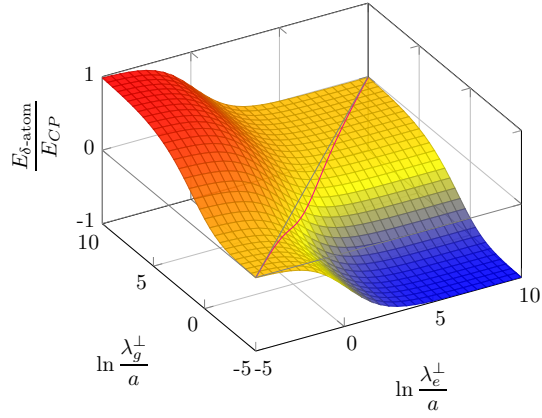}
\caption{The variation of the Casimir-Polder interaction energy between an 
isotropic atom and a magneto-electric $\delta$-function plate in units of the 
magnitude of the usual Casimir-Polder interaction energy between an isotropic 
atom and a perfect electrically conducting plate as a function of 
$\ln\frac{\lambda^\perp_e}{a}$ and $\ln\frac{\lambda^\perp_g}{a}$.}
\label{cp-me}
\end{center}
\end{figure}

In Fig.~\ref{cp-me} we show the variation of the Casimir-Polder energy given in 
Eq.~(\ref{CP12-tp}) with respect to the electric and magnetic properties of the 
magneto-electric $\delta$-function plate in units of the distance $a$ between the 
plates. We set $\alpha^\perp=\alpha^{||}$. The energy is normalized relative to 
the magnitude of the usual Casimir-Polder energy for an isotropic atom 
interacting with a perfect electric conductor. It is of interest to note that the 
interaction energy is always negative when the plate is purely electric and 
always positive when the plate is purely magnetic. The transition from a negative 
to a positive value of the energy occurs along a curve in the 
$\lambda^\perp_e$-$\lambda^\perp_g$ parameter space. 
In particular, the interaction energy 
vanishes for
\begin{subequations}
\begin{eqnarray}
\frac{\lambda^\perp_e}{a} &=& \frac{3}{7} \frac{\lambda^\perp_g}{a}
\qquad \qquad \qquad \text{for the weak coupling limit} 
\left(\lambda^\perp_{e,g}\ll a\right), \label{weak}\\
\frac{\lambda^\perp_e}{a} &=& \frac{256}{45} \frac{1}{\pi^{3/2}}
\sqrt{\frac{\lambda^\perp_g}{a}}
\qquad \text{for the strong coupling limit} 
\left(\lambda^\perp_{e,g}\gg a\right),\label{strong}%
\end{eqnarray}%
\end{subequations}%
to the leading order. Interestingly for the strong coupling case the interaction 
energy scales differently for the magnetic coupling $\lambda^\perp_g$ as compared 
to the electric coupling $\lambda^\perp_e$. Furthermore, the force between an 
isotropic atom and a magneto-electric $\delta$-function plate changes sign for 
different combination of $\lambda^\perp_e$ and $\lambda^\perp_g$. For example, 
for strong coupling the force vanishes for a condition of the form 
Eq.~(\ref{strong}) where numerical coefficient is replaced by $1536/315$.

The total reflection coefficients, $r^H$ and $r^E$, vanish for the special 
case when the plate behaves like a perfect magneto-electric conductor, i.e., 
$\lambda^\perp_e \to \infty$ and $\lambda^\perp_g \to \infty$. Thus, the 
Casimir-Polder interaction energy also vanishes for such a plate. This is a 
generic behavior for a perfectly conducting magneto-electric $\delta$-function 
plate. For a perfect electric conductor the TM and TE reflection coefficients are 
$r^H=1$ and $r^E=-1$ in which case we obtain the usual Casimir-Polder energy 
between an atom and a perfect electric conductor. In contrast, for a perfect 
magnetic conductor $r^H=-1$ and $r^E=1$ we obtain a repulsive interaction energy 
of the same magnitude, as evident from Fig.~\ref{cp-me}.
\subsection{Atom interacting with a magneto-electric $\delta$-function 
plate on a dielectric substrate}

\begin{figure}
\subfigure[Small electric permittivity]
{\includegraphics[width=65mm]{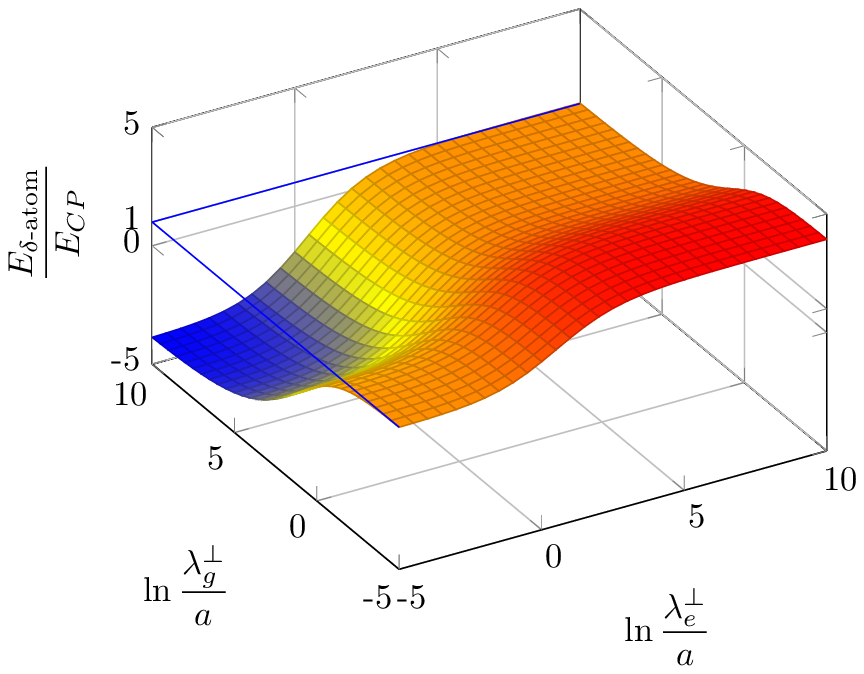}
\label{ep2-fig}}
\hspace{3mm}
\subfigure[High electric permittivity] 
{\includegraphics[width=65mm]{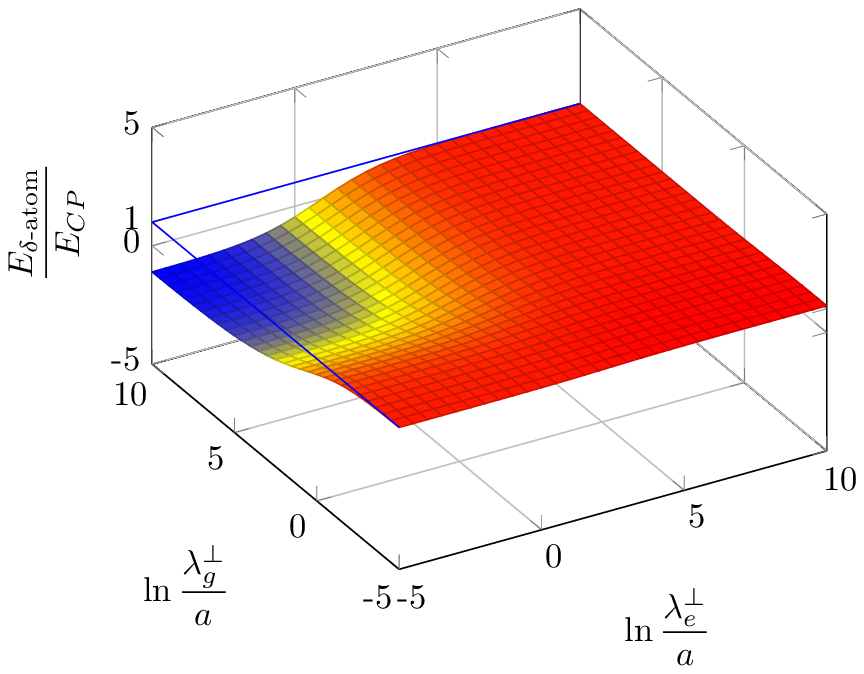}
\label{ep100-fig}}
\caption{Fractional change in the Casimir-Polder energy between 
an isotropic atom and a dielectric substrate of a fixed electric 
permittivity in the presence of the magneto-electric $\delta$-function plate 
relative to the Casimir-Polder energy in the absence of the plate as a function 
of the magnetic and electric properties of the plate, 
$\ln\frac{\lambda^\perp_e}{a}$ and $\ln\frac{\lambda^\perp_g}{a}$, respectively. 
The electric permittivity of the substrate material is \subref{ep2-fig} 
$\varepsilon=2$ and \subref{ep100-fig} $\varepsilon=100$.}
\end{figure}

As a second example let us consider an anisotropic atom interacting with a 
magneto-electric $\delta$-function plate on a semi-infinite dielectric substrate 
as shown in Fig.~\ref{atom-plate-slab}. The Casimir-Polder energy is still 
expressed by Eq.~(\ref{CP12-tp}) with the reflection coefficients, $r^H$ and 
$r^E$, now obtained from Eq.~(\ref{rH-tH}). We choose the semi-infinite material 
to be isotropic and non-magnetic to reduce the numbers of parameters in the 
analysis. 
Again we set $\alpha^\perp=\alpha^{||}$ for the atom. Figures~\ref{ep2-fig} and 
\ref{ep100-fig} show the fractional change in the Casimir-Polder energy in the 
presence of a magneto-electric $\delta$-function plate compared to the absence 
of the magneto-electric $\delta$-function plate on the substrate. When the 
electric permittivity of the substrate material is low then the presence of the 
magneto-electric $\delta$-function plate increases the magnitude of the 
interaction energy depending on the material properties of the plate, while the 
variation is less strong in the case when the dielectric permittivity of 
the substrate material is high. The biggest effect occurs when $\lambda_g$ is 
large and $\lambda_e$ is small. In other words, the material with stronger 
properties dominates in the contribution to the interaction energy.
%
\section{Interaction energy between two magneto-electric $\delta$-function 
plates}
\begin{figure}\begin{center}
\includegraphics[width=5cm]{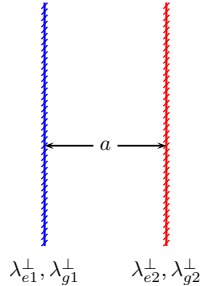}
\label{c-d-p-fig}
\caption{Parallel magneto-electric $\delta$-function plates separated by a 
distance $a$.}
\end{center}\end{figure}

In this section we evaluate the Casimir interaction energy between two 
magneto-electric $\delta$-function plates and study its variation as a function 
of the electric and magnetic properties of the plates.
Let us consider two $\delta$-function plates described by the electric and 
magnetic properties, $\lambda^\perp_{ei}$ and $\lambda^\perp_{gi}$, respectively, 
with subscript $i=1,2$ representing the individual plates. The separation 
distance between the plates is $a$. See Fig.~\ref{c-d-p-fig}. 
Considering that the TM and TE modes decouple for the planar geometry, the 
Casimir energy is conveniently expressed as
%
\begin{equation}
\frac{E_\text{$\delta$-plate}}{A}
= \frac{1}{2} \int_{-\infty}^\infty \frac{d\zeta}{2\pi}
\int \frac{d^2k_\perp}{(2\pi)^2} \Bigg\{ \ln \Big[ 1
- r_1^H r_2^H\, e^{-2\kappa a}\Big] + \ln \Big[ 1
- r_1^E r_2^E \,e^{-2\kappa a}\Big] \Bigg\},
\label{E12tp-Cas}
\end{equation}
%
where the TM reflection coefficient for a single magneto-electric 
$\delta$-function plate is given in Eq.~(\ref{rH-tot}). The TE reflection 
coefficient $r^E$ is obtained by replacing $e \leftrightarrow g$ and $H \to E$. 
As mentioned before, the interaction energy vanishes when both plates are 
perfect magneto-electric conductors, as if plates are invisible to each other. 

\begin{figure}
\subfigure[Magneto-electric $\delta$-function plates with 
$\lambda^\perp_{e1}=\lambda^\perp_{e2}$ and 
$\lambda^\perp_{g1}=\lambda^\perp_{g2}$.]
{\includegraphics[width=6cm]{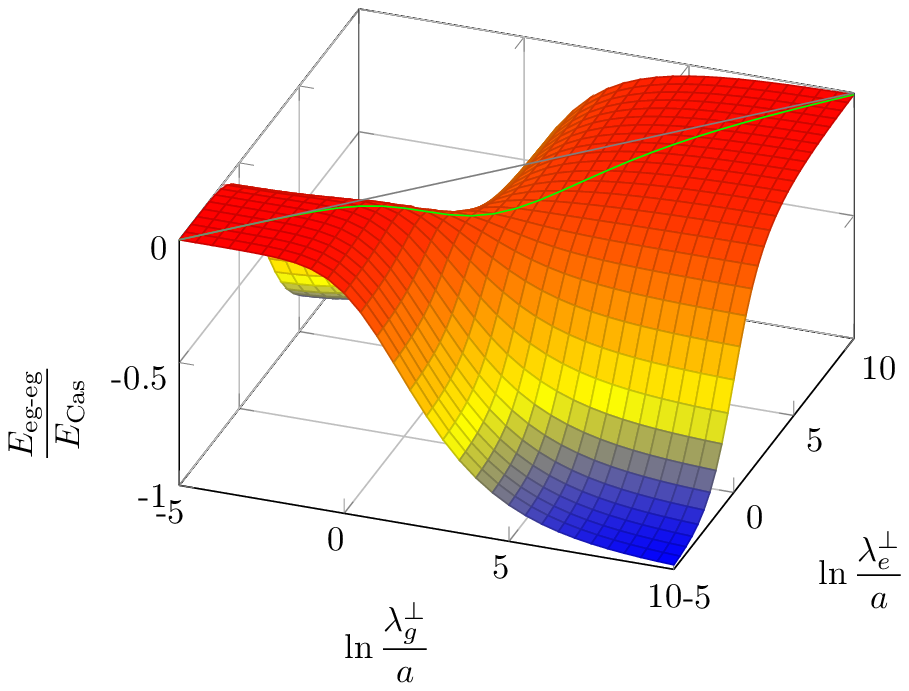}
\label{egeg} }
\hspace{10mm}
\subfigure[Magneto-electric $\delta$-function plates with 
$\lambda^\perp_{e1}=\lambda^\perp_{g2}$ 
and $\lambda^\perp_{g1}=\lambda^\perp_{e2}$.]
{\includegraphics[width=6cm]{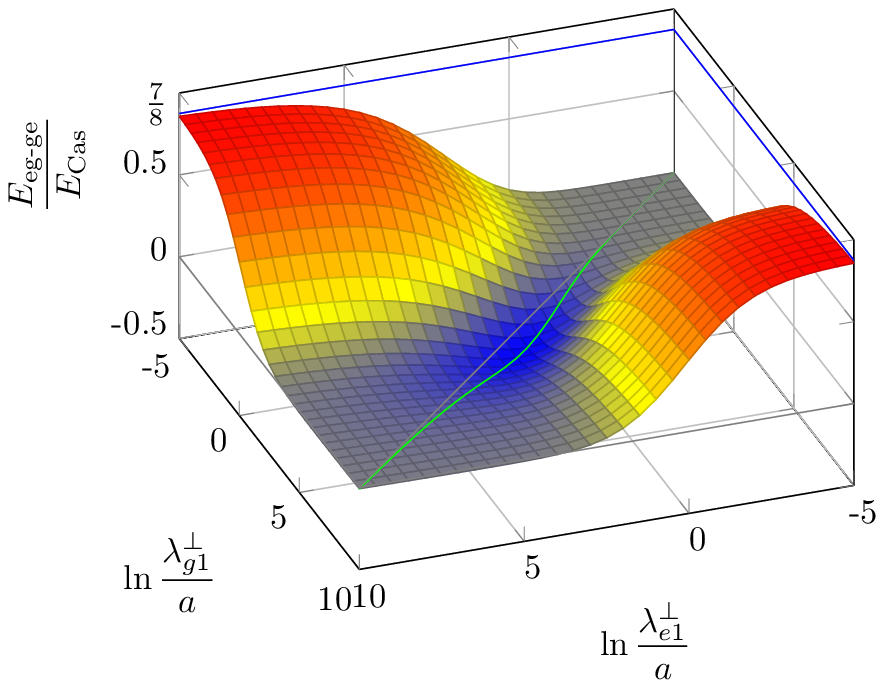}
\label{egge} }
\caption{Fractional change in the Casimir energy of parallel magneto-electric 
$\delta$-function plates separated by distance $a$ relative to the magnitude of 
the Casimir energy of two perfect electrically conducting parallel plates as a 
function of the electric and magnetic properties of the $\delta$-function plate, 
$\ln\frac{\lambda^\perp_e}{a}$ and $\ln\frac{\lambda^\perp_g}{a}$, respectively. 
In \subref{egeg} we assume the magnetic and electric properties of two plates to 
be same. The interaction energy vanishes when both 
$\frac{\lambda^\perp_e}{a}\rightarrow\infty$ and 
$\frac{\lambda^\perp_g}{a}\rightarrow\infty$. In \subref{egge} we assume the two 
plates have dual properties, i.e., $\lambda^\perp_{e1}=\lambda^\perp_{g2}$ and 
$\lambda^\perp_{g1}=\lambda^\perp_{e2}$. The ratio of energies approaches Boyer's 
result for the Casimir energy between a perfect electrically conducting plate and 
a perfect magnetically conducting plate. 
}
\end{figure}

In Fig.~\ref{egeg} we plot the ratio of the Casimir interaction energy given in 
Eq.~(\ref{E12tp-Cas}) to the magnitude of the Casimir energy between two 
perfectly conducting electric plates as a function of the electric and magnetic 
properties. For simplicity we have set 
$\lambda^\perp_{(e,g)1}=\lambda^\perp_{(e,g)2}$. The fractional change in the 
energy vanishes when there are no plates and when both the plates are 
simultaneously perfect electric and perfect magnetic conductors. In the case when 
both plates are either perfect electric conductors or perfect magnetic 
conductors, the energy ratio approaches $-1$ as expected. The ratio of the 
energies is always negative except when it goes to zero for two extreme cases 
described above. In addition, it is easy to check that the force between two 
identical magneto-electric $\delta$-function plates is always attractive by 
taking a negative derivative of Eq.~(\ref{E12tp-Cas}) with respect to the 
separation distance $a$. Kenneth and Klich 
in~\cite{Kenneth-Klich-PhysRevLett.97.160401} proved that for non-magnetic bodies 
``the Casimir force between two bodies related by reflection is always 
attractive, independent of the exact form of the bodies or dielectric 
properties''. The above example is a generalization of their theorem to 
magneto-electric bodies. The magnitude of the interaction energy, in general, is 
less than the usual Casimir energy between two perfect electrically conducting 
plates. The green line on the energy surface in Fig.~\ref{egeg} shows the value 
of the ratio of the interaction energies in the case 
$\lambda^\perp_e=\lambda^\perp_g$.

We plot another interesting case in Fig.~\ref{egge}, where we have considered the 
material properties of the two plates to be dual of each other, i.e. 
$\lambda^\perp_{e1}=\lambda^\perp_{g2}$ and 
$\lambda^\perp_{g1}=\lambda^\perp_{e2}$. The interaction energy vanishes for the 
two cases when both the plate properties vanish, i.e. no plates, or both approach 
the perfect magneto-electric conductor limit where the plates become transparent 
to electromagnetic fields. In addition, the interaction energy in this case can 
be either negative, positive, or zero, the latter occurring for a specific 
combination of values of $\lambda^\perp_{e1}$ and $\lambda^\perp_{g1}$. The green 
line on the energy surface in Fig.~\ref{egge} shows the value of the ratio of the 
interaction energies when 
$\lambda^\perp_{e1}=\lambda^\perp_{g2}=\lambda^\perp_{g1}=\lambda^\perp_{e2}$. 
The interaction energy approaches Boyer's result~\cite{Boyer-PhysRevA.9.2078} for 
the Casimir energy between a perfect electrically conducting plate and a perfect 
magnetically conducting plate when 
$\lambda^\perp_{e1}=\lambda^\perp_{g2} \to \infty$ and 
$\lambda^\perp_{e2}=\lambda^\perp_{g1} \to 0$ or {\it vice versa}:
\begin{equation}
E_{e\text{-}g}=+\frac{7}{8}\frac{\pi^2}{720 a^3}. 
\label{Boyer}
\end{equation}

\section{Conclusions}
In this paper we have extended our investigation of the magneto-electric 
$\delta$-function plates initiated in~\cite{Parashar-PhysRevD.86.085021}. A 
$\delta$-function plate having both electric and magnetic properties has an 
interesting property of optically vanishing in the simultaneous perfect electric 
and perfect magnetic conducting limit, which is a generic property. It can be 
physically realized in nature by a plasma slab of thickness $d$ in the thin-plate 
limit, where the characteristic wavenumber $\zeta_p=\omega_p^2 d$ satisfies the 
constraint: $\zeta d \ll \sqrt{\zeta_p d}\ll 1$. The Casimir-Polder energy of 
such a plate with an isotropic atom is always negative when the plate is purely 
electric and always positive when the plate is purely magnetic. The presence of 
a magneto-electric $\delta$-function plate on a dielectric medium changes the 
Casimir-Polder energy by shielding the medium with significant variation observed 
when the medium is weakly interacting. For the case of interaction between two 
identical $\delta$-function plates we find that the force is always attractive 
and vanishes when the plates become simultaneously perfect electric and perfect 
magnetic conductors. However, if the two $\delta$-function plates have dual 
properties, i.e., the electric and magnetic properties of one plate are 
interchanged in the second plate, then the plates can either attract, repel, or 
experience vanishing force, where latter occurs for a specific set of values of 
the electric and magnetic properties. It approaches Boyer's result when 
one plate becomes a perfect electric conductor and the other plate becomes a 
perfect magnetic conductor.

\acknowledgments
KAM and PP would like to acknowledge the financial support from the US National 
Science Foundation Grant, No.~0968492, and the Julian Schwinger Foundation. MS 
would like to acknowledge support by US National Science Foundation Grant 
No.~PHY-09-02054. We thank Ryan Behunin, Cynthia Reichhardt, Elom Abalo, Fardin 
Kherandish, and Baris Altunkaynak for discussions. 

\bibliographystyle{varenna}
\bibliography{biblio/MSQS-paper}

\end{document}